\documentclass[12pt]{article}
\usepackage{epsf}
\usepackage{epsfig}

\def\be{\begin{equation}}
\def\ee{\end{equation}}
\def\bea{\begin{eqnarray}}
\def\eea{\end{eqnarray}}

\newcommand{\bee}{\begin{eqnarray}}
\newcommand{\eee}{\end{eqnarray}}

\begin{document}


\begin{center}
\section*{Kaon Electromagnetic Form Factor in the Light-Front
Formalism\protect
\footnote{To {\it appear ``Physics of Elementary Particles and 
Atomic Nuclei, Vol.~{\bf 36}, (2005).}''}}

\vskip 5mm
Fabiano P. Pereira$^{a}$, J.P.B.C. de Melo$^{a,,b}$,
T. Frederico$^{c}$ and Lauro Tomio$^{a}$
\vskip 5mm

{\small
(a) {\it Instituto de F\'\i sica Te\'orica, Universidade Estadual Paulista,\\
01405-900, S\~ao Paulo, SP, Brazil}
\\
(b) {\it Centro de Ci\^encias Exatas e Tecnol\'ogicas, Universidade
Cruzeiro do Sul,\\ 08060-070, S\~ao Paulo, Brazil}
\\
(c) {\it Departamento de F\'{\i}sica, ITA, Centro T\'ecnico
Aeroespacial,\\ 12228-900, S\~ao Jos\'e dos Campos, Brazil}
}
\end{center}
\vskip 5mm

\begin{abstract}
Numerical calculations are performed and compared to the
experimental data for the electromagnetic form factor of the
kaon, which is extracted from both components of the electromagnetic
current, $J^{+}$ and $J^{-}$, with a pseudo-scalar coupling of
the quarks to the kaon. In the case of $J^{+}$ there is no pair term
contribution in the Drell-Yan frame ($q^{+}=0$). However, for
$J^{-}$, the pair term contribution is different from zero and
necessary in order to preserve the rotational symmetry of the current.
The free parameters are the quark masses and the regulator mass.
\end{abstract}
\vskip 8mm

\subsection*{1. Introduction}

The convenience of using light-front variables in QCD descriptions
of hadron properties and interactions, has been established long
time ago. In particular, we have a clarifying article by
N.N. Bogolyubov and coworkers published in 1983~\cite{bogo}.
In the subsections 3.5.3 and 3.5.4 of this article, the QCD
description of the simplest composite systems (the mesons) and the
corresponding form factors at high momentum transfer are discussed.
The formalism is developed from a gauge invariant two-point function
of the Bethe-Salpeter amplitude. The pion electromagnetic form
factors are shown as example using the light-front
formalism.
Models for wave-functions in the light front are originally
developed in \cite{teren}.
In more recent years, the use of light-front formalism has become a
common procedure in QCD description of hadrons~\cite{pr-BPP}.
Here, we can mention a few works that we are concerned, as 
\cite{Chung88,Ji90,Frederico92,Cardarelli96,doro97,Pacheco97,
Pacheco2002}, dedicated to
study pseudoscalar properties of mesons, structure wave functions and
quark-antiquark correlations. From such references, one can trace a
more complete and detailed bibliography.

In the present communication, we report results for the kaon electromagnetic
form factor that are extracted from both components of the electromagnetic
current, $J^{+}=J^0 + J^3$ and $J^{-}=J^0-J^3$, with a pseudo-scalar
coupling of the quarks.
In the case of $J^{+}$ there is no pair term contribution in the Drell-Yan
frame ($q^{+}=0$). However, for the $J^{-}$ component of the electromagnetic
current,  the pair term contribution is
different from zero and necessary in order to preserve
the rotational symmetry of the current.
We note that, when considering the case of vector particles, even the
$J^+$ electromagnetic current has contribution from pair terms
to have a full covariant theory  \cite{Pacheco98,Jaus99}
(For a more recent application of this ideas in the vector anomaly problem, 
see ref.\cite{bf05}).
In order to satisfy the angular condition for spin one particles, it is
necessary to consider  pair terms in the electromagnetic current
$J^+$~\cite{Pacheco98}.
 Besides the valence contribution to the $J^-$ current, the pair term is
necessary  for both, pseudoscalar and vector particles to keep the rotational
symmetry properties of the current in the light-front formalism.

\section*{2. Electromagnetic Current Model}

In order to extract the electromagnetic form factor for the kaon, the
components $J^{+}$ and $J^{-}$ of the electromagnetic current are used.
The $J^{(\mu=\pm)}$ components of the electromagnetic current for the
kaon have contribution, from the quark ($q$) and the antiquark ($\bar q$),
are given by
\begin{eqnarray}
J_{q}^\mu (q^2)=&&\imath e_{q} g^2 N_c \int
\frac{d^4k}{(2\pi)^4}\times \nonumber \\
\times&&{\rm Tr}[S(k,m_{\bar{q}})
\gamma^5 S((k-P^{\prime}),m_{q}) \gamma^\mu S((k-P),m_{q})
\gamma^5 ] \Lambda(k,P^{\prime})
\Lambda(k,P) \ ,  \label{j+kaon} \nonumber \\
J^\mu_{\bar{q}}(q^2)=&& q \leftrightarrow \bar {q} \ \
\mbox{in} \ \ J^{\mu}_q(q^2),
\end{eqnarray}
where the number of colors is $N_c=3$, $g$ is the coupling constant
and $e_q$ ($e_{\bar q}$) is the quark (anti-quark) charge.
We use the Breit frame, where the momentum transfer is
$q^2=-(\vec q_\perp)^2$, $P^0=P^{\prime \, 0}$ and
$\vec P^{\prime}_\perp=-\vec P_\perp=\vec{q_\perp}/{2}$.
The function $\Lambda(k,p)=N/[(p-k)^2-m^2_{R}+\imath \epsilon]$
is used in order to regulate the divergent integral, where
$m_R$ is the regulator mass and $m_{q}$ and $m_{\bar q}$
are, respectively, the quark and anti-quark masses.
The function $S(p)$ is the fermion propagator:
\begin{equation}
\displaystyle
S(p,m)=\frac{1}{\rlap\slash p-m+\imath \epsilon}.
\end{equation}

The light-front coordinates are defined as
$k^+=k^0+k^3 \ , k^-=k^0-k^3 \ , k_\perp=(k^1,k^2)\,$.
In the following, for the calculation of the pair terms,
we consider the model given in \cite{Pacheco98} for a composite
boson bound state and in the study of the Ward-Takahashi identity
in the light-front formalism \cite{Naus98}.
The contribution of the pair term for $J^{+}$ and $J^{-}$ components
of the electromagnetic current comes from the matrix elements
proportional to $k^-$ in both cases (antiquark and quark on-shell).

\section*{3. Electromagnetic Form Factor}

The most general expression for the form factor of the spin zero
particles is given by:
\begin{equation}
\langle P|J^{\mu}|P^{\prime}\rangle = (P^{\prime} + P)^{\mu} F(q^2) +
(P^{\prime}-P)^{\mu} G(q^2)
\end{equation}
In this elastic process, the form factor depends only on $q^2$,
and $G(q^2)=0$ in all $q^2$. Here, off-shell effects are
not explored. However, the off-shell effects are important
and relevant in many topics for particles and nuclear physics.

In order to extract the form factor for the kaon,
$F_{K^+}(q^2)$, we used both $J^{+}$ and
$J^{-}$ components of the electromagnetic current.
One can verify that only the on-shell pole
$\displaystyle \bar{k}^{-}=
(k_{\perp}^{2}+m^2_{\bar q})/k^+$
contribute to the $k^{-}$ integration in the
interval $0 < k^{+} <  P^{+}$:
\begin{eqnarray}
F^{+}_{\bar q}(q^2)  &=&   -e_{\bar q} \frac{N^2 g^2 N_c}{P^{+}}
\int \frac{d^{2} k_{\perp} d k^{+} }{4 \pi^3}  
\frac{ {\cal N}^+_{\bar q} \;\; \theta(k^{+})\theta(P^{+}-k^{+})} 
{k^+(P^+-k^+)^2 (P^{^{\prime}+}-k^+)^2(P^- -
\bar{k}^- - \frac{f_{2,q} -\imath \epsilon}{P^+ - k^+})
}   \nonumber \\
&&\times  \frac{1} { (P^{\prime -} - \bar{k}^{-}  - \frac{f_{3,q}
-\imath \epsilon }{P^{\prime +} - k^+}) (P^- - \bar{k}^{-}  -
\frac{f_4 -\imath \epsilon }{P^+ - k^+}) (P^{\prime -} - \bar{k}^{-}
-
\frac{f_5 -\imath \epsilon }{P^{\prime +} - k^+})} , \label{ff4} \\
\ F^{+}_{q}(q^2) & = & [ \ q \leftrightarrow \bar{q} \ \  \mbox{in}
\ F^{+}_{\bar q}(q^2) \ ]   \  ,
\label{ffactor+}
\end{eqnarray}
where $f_{2,\;q}\equiv (P - k)_{\perp}^{2}+m^2_{ q}$, $f_{3,\;q}\equiv(P^{\prime} - k)_{\perp}^{2}+m^2_{ q}$, 
$f_4\equiv(P-k)_{\perp}^{2}+m^2_{R}$ and
$f_5\equiv(P^{\prime}-k)_{\perp}^{2}+m^2_{R}$ ~. In the numerator,
${\cal N}^+_{\bar q}$ is given by
\begin{eqnarray}
&&{\cal N}^+_{\bar q}=\left. \frac{-1}{4}{\rm Tr}[(\rlap\slash k +m_{\bar{q}})
\gamma^5 (\rlap\slash k-\rlap\slash P^{\prime}+m_{q}) 
\gamma^+ (\rlap\slash k-\rlap\slash P +m_{q})
\gamma^5 ]\right|_{k_-=\bar{k}^{-}}
.\end{eqnarray}
The kaon light-front wave function of the model can be extracted
from (\ref{ff4})and (\ref{ffactor+}) as
\begin{eqnarray}
\Phi^i_{Q}(x,k_\perp)=\frac{1}{(1-x)^2}
\frac{N}{(m^2_{K^+}-M^2_{0}) (m^2_{K^+}-M^2(m_{Q},m_R))} \ ,\label{fupi}
\end{eqnarray}
where $x=k^+/P^+$ is the momentum fraction, $Q= {\bar q}, q$ and 
\begin{equation}
M^2(m_Q,m_R)=\frac{k^2_\perp+m^2_Q}{x}+
\frac{(P-k)^2_\perp+m^2_R}{1-x}-P^2_\perp 
\label{fupi2} .
\end{equation}
The squared free quark mass is given by $M^2_{0}=M^2(m_{\bar
q},m_q)$. For the final wave-functions, $ \Phi^f_{\bar q}$ and $\Phi^f_{q}$, 
we just need to exchange $P \leftrightarrow P^{\prime  }$ in (\ref{fupi}) and
(\ref{fupi2}).

The expression obtained for the electromagnetic form factor in terms
of the initial $(\Phi^i_{\bar q})$ and final
$(\Phi^f_{\bar q})$ wave functions is
{\small
\begin{eqnarray}
F^{+}_{\bar q}(q^2)&=& -e_{\bar  q} \frac{N^2 g^2 N_c}{4\pi^3 P^+}
\int \frac{d^{2} k_{\perp} d x} {x}  {\cal N}^+_{\bar q}\theta(x)
\theta(1-x) \ \Phi^{*f}_{\bar q}(x,k_{\perp})
\Phi^i_{\bar q}(x,k_{\perp})  \ , \\
\ F^{+}_{q}(q^2) & = & [\ q \ \leftrightarrow \ \bar{q}  \  \mbox{in}
\ F^{+}_{\bar q}(q^2) \ ] \  \ .
\label{form}
\end{eqnarray}
 }

\begin{figure}[h]
\vspace{5cm}
\includegraphics{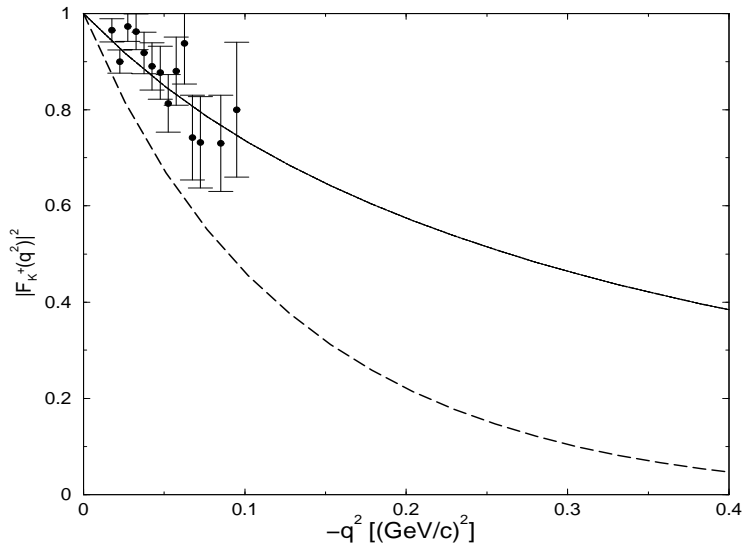}
\vskip 1.5cm
\caption{
Kaon ($K^+$) electromagnetic form factor calculated
within the covariant and the light-front formalisms, using
$J^{+}$ and $J^{-}$ components of the electromagnetic current.
The dashed line give the results from $J^{-}$ without the
light-front pair term. Adding the pair term to it, we obtain
the result given by the solid line, which coincides with
the light-front and covariant calculations with $J^{+}$.
Experimental data are from \cite{Amendolia86}.
}
\label{fig1}
\end{figure}

The final expression for the
electromagnetic form factor obtained with $J^{+}$  is the sum of
two contributions from the quark and the antiquark currents:
\begin{equation}
F_{K^+}^{+}(q^2)=F_{q}^{+}(q^2)+F_{\bar{q}}^{+}(q^2) \ ,
\end{equation}
where the normalization is given by $F^{+}_{K^+}(0)=1$.
The calculation of the kaon electromagnetic form factor
in the light-front with  $J^+$, without pair term, gives the same
result as the covariant one (see Fig.\ref{fig1}).

The contribution to the electromagnetic form factor obtained with
$J^{-}$ after the integration in $k^{-}$ from the interval $0 < k^+
< P^{+}$ is given by
\begin{eqnarray}
F^{- (I)}_{\bar q}(q^2) & = & -e_{\bar q} \frac{N^2 g^2 N_c }
{4 \pi^3 P^{+}}
\int \frac{d^{2} k_{\perp} d x}{ x } \theta(x)
\theta(1-x) {\cal N}^{-(I)}_{\bar q}  \ \Phi^{*f}_{\bar
q}(x,k_{\perp})
\Phi_{\bar q}^{i}(x,k_{\perp}), \\
\ F^{-(I)}_{q}(q^2) & = & [ \ q \ \leftrightarrow \ \bar{q}  \
\mbox{in} \ F^{-(I)}_{\bar q}(q^2) \ ] , \label{final2-}
\end{eqnarray}
where 
\begin{eqnarray}
{\cal N}^{-(I)}_{\bar q} &=&
\frac{k_{\perp}^{2}+m^2_{\bar q}}{x P^+}\left[(m_{q}-m_{\bar{q}})^2-\frac{q^2}{4}\right]+
P^+\left[2 m_{\bar{q}}(m_{q}-m_{\bar{q}})+ x P^{+ 2}\right]
\end{eqnarray}

When using  $J^-$ to extract the electromagnetic form factor,
besides the contribution of the interval (I), the pair term
contributes to the  electromagnetic form factor in the interval (II)
($P^+ < k^+  < P^{\prime +}$). The pair term contribution for the
form factor, as shown in \cite{Pacheco99,Pacheco2002}, is given by
$F^{-(II)}(q^2)$.
\begin{equation}
F^{-(II)}(q^2)= \frac{N^2 g^2 N_c}{P^+} \bigg[ e_{q}
\Delta_{q}^{-(II)}(q^2) + e_{\bar{q}}\Delta_{\bar{q}}^{-(II)}(q^2)
\bigg]
\end{equation}
where $\Delta_{\bar{q}}^{-(II)}$ and $\Delta_{q}^{-(II)}$  given
below. These terms correspond to the pair contribution in the
$J^{-}$ component of the electromagnetic current, which are obtained
after the integration in $k^-$ and the limit $P^{\prime
+}\rightarrow P^+$ are performed. Then one gets the following
equations for the pair terms:
\begin{eqnarray}
\Delta_{\bar{q}}^{-(II)}= & & \frac{-1}{4 \pi^3}\int \frac{d^2 k_{\perp}}{(2
\pi)^4} {\cal N}^{-(II)}_{\bar q}   \sum_{i=2}^{5}\frac{ \ln(f_{i})}
{\prod_{j=2,i\neq j}^{5}(f_j-f_i)}  \ , \label{par-1}
\end{eqnarray}
where $f_2\equiv f_{2,\;q}$, $f_3\equiv f_{3,\;q}$ and 
\begin{eqnarray}
{\cal N}^{-(II)}_{\bar q}= \frac{-1}{P^+}\left[
P^{+2} + \frac{q^2}{4} - (m_{q}-m_{\bar{q}})^2\right] \label{nqbar}.
\end{eqnarray}
We obtain the corresponding quark current contribution as in the
above Eqs.  (\ref{par-1}) and (\ref{nqbar}) just by replacing 
$q \leftrightarrow \bar{q}$.

In the limit $P^{\prime +}\rightarrow P^+$, the pair term
contribution (zero mode) is non zero and responsible for the
covariance of the $J^{-}$ component of the electromagnetic current.
The sum of the contributions from the intervals (I) and (II) for
$J^{-}$ in the light-front gives the same result as in the 
covariant calculation~\cite{Pacheco98,Naus98}.

The final expression for the electromagnetic form factor for the
kaon, extracted from $J^{-}$ is
\begin{equation}
F_{K^+}^{-}(q^2)=\big[F^{-(I)}_{q}(q^2) + F^{-(I)}_{\bar{q}}(q^2) +
F^{-(II)}(q^2)
\big]
\label{final-} \ ,
\end{equation}
which is normalized by the charge conservation, $F_{K^+}^{-}(0)=1$.

\section*{4. Results and Conclusion}

Next, we present results obtained considering the light-front
formalism, as well as the covariant formalism.
The parameters of the model are the constituent quark masses
$m_q=m_{u}=0.220$ GeV, $m_{\bar{q}}=m_{\bar{s}}=0.419$ GeV, and
the regulator mass $m_R=0.946$ GeV, which are adjusted to fit
the electromagnetic radius of the kaon.
With these parameters, the calculated electromagnetic
radius of the kaon is $\langle r^2_{k^+}\rangle =0.354$ fm$^2$,
very close to the experimental radius $\langle r^2_{k^+}\rangle
=0.340$ fm$^2$ \cite{Amendolia86}.

The electromagnetic form factor is presented in Fig.\ref{fig1}.
Due to the fact that $J^{+}$ does not
have light-front pair term contributions, the electromagnetic form
factor results equal to the one obtained in a covariant calculation.
In the case of  $J^{-}$, the light-front calculation gives results
quite different from the covariant results, as shown in
figure \ref{fig1}.
After the inclusion of the pair terms, with $J^{-}$, we observe a 
complete agreement between the light-front and covariant 
results.
In conclusion, the $J^{+}$ and $J^{-}$ components of the
electromagnetic current of the kaon are obtained in the light-front
and in the covariant formalisms, in a constituent quark model.
In the case of  $J^{-}$, we note that the pair terms are essential to 
obtain a complete agreement between the covariant and the light-front 
results for the kaon electromagnetic form factor.

Our thanks to the Brazilian agencies FAPESP and CNPq for partial 
support.   \\

\end{document}